# Two electrons in the Dirac-bubble potential well

Running head: Two electrons in $C_{60}$ potential well

A S Baltenkov
Institute of Ion-Plasma and Laser Technologies, Uzbek Academy of Sciences
100125, Tashkent, Uzbekistan, arkbalt@mail.ru

**Abstract.** Within the Dirac-bubble potential model an electronic structure of the double-charged negative ion $C_{60}^{2-}$ has been studied by a variational method. It has been shown that even in the first approximation of this method when a trial wave function of two electrons is represented as a product of one-electron functions the total energy of system is negative, which is evidence of a stable state of the double-charged negative ion in this model. The photodetachment cross sections of this ion have been calculated. It has been shown that the threshold behavior of the cross sections is different. The first cross section accompanying by transformation of double-charged ion into one-charged one is exponentially small near the process threshold. The second cross section – one-charged ion photodetachment - increases at the threshold as a power function. The values of these cross sections are of the same order as the photodetachment cross sections of atomic negative ions with the same affinity energy.

## 1. Introduction

The photoionization of atom A in the endohedral anions: $A@C_{60}^{z-}$ was discussed in Refs. [1,2]. The charge of the negative molecular ion $z$ was varied within the range $0 \le z \le 5$ in those calculations. The question whether the many-charged negative ions of $C_{60}$ inside the hollow interior of which an atom A can be encapsulated did not discussed and the existence of the $A@C_{60}^{z-}$ systems was simply postulated in [1,2]. Atomic multi-charged ions (even for $z$=2) are, as a rule, unstable systems with the lifetimes $10^{-6}$–$10^{-7}$ s [3]. Their instability is due to Coulomb repulsion of extra electrons the wave functions of which have the parent atom nucleus as a center. Because of strong overlapping of electron densities the energy of Coulomb interaction prevails over the binding energy of electron with atom, leading to system decay. It can be expected that conditions of existence of multi-charged anions $C_{60}^{z-}$ are more favorable. It is easy to imagine that extra electrons (for example, for $z$=2) are localized mainly at the opposite sides of the fullerene sphere to minimize the Coulomb repulsion. Since the sphere radius is big in atomic scale the repulsion energy can prove to be insufficient for decay. The energy of $z$-electrons locked in the short-range potential well $U(r)$ of the $C_{60}$ shell and forming the multi-charged ion $C_{60}^{z-}$ is negative: $E_0 < 0$; while the energy of Coulomb repulsive of these electrons is positive: $Q > 0$. A stable bound state of the electrons in the potential well is possible if their total energy is $E_0 + Q < 0$. In the first approximation the energies $E_0$ and $Q$ can be estimated as follows. The binding energy of one electron with $C_{60}$ is $\approx -2.7$ eV [4]. The upper estimate for the energy of $z$ electrons in the well is $E_0 \approx -2.7 \times z$ eV *. The Coulomb repulsion energy of the pair of electrons located in the potential well of the radius $R$=3.55 Å [4] has the value of about $e^2/R \approx 4.1$ eV. A number of such pairs is a binion $p = z!/[2(z-2)!]$. Now we have the following estimation for the total energy of electrons: $E_0 + Q \approx -2.7z + 4.1p$. Substituting $z$ = 5, 4 and 3 in this formula, we come to conclusion that the total energy of penta-, tetra- and three-charged negative ions of $C_{60}$ is positive and so it is unlikely they can exist in a stable state. The total energy is negative for $z$=2 only.

In connection with that it is interesting to analyze whether the existence of stable double-charged negative ions of $C_{60}$ is possible, in principle, and if so what their electron structure is. This paper is devoted to this question. The energy state of a system consisting of two electrons moving in the Dirac-bubble potential well is studied. We will assume [5] that in the $1s^2$ ground state of the system the both electrons are in states with zero orbital moment and antiparallel spins. The system in question is similar to a helium-like positive ion with the only difference that the coupling of each electron of the system is due to the short-range potential $U(r)$ rather than the long-range interaction of electrons with the positive ion nucleus. Therefore, to analyze this problem we will apply the methods used for helium-like positive ions [6,7].

First, the energy of the $C_{60}^{2-}$ ground state will be calculated considering the Coulomb interaction between electrons as a perturbation (Sec. 2). In Section 3 with the help of variational principle we will calculate the binding energy of the anion ground state and corresponding wave functions within the framework of the Dirac-bubble potential model. In Sec. 4 the wave functions will be used to calculate the photodetachment cross sections of the following processes $C_{60}^{2-} + \hbar\omega = C_{60}^{-} + e$ and $C_{60}^{-} + \hbar\omega = C_{60} + e$ ($\hbar\omega$ is the photon energy). Section 5 gives conclusions.

**2. Mean value of energy of Coulomb interaction**

Let us study the behavior of a pair of the electrons in the spherically symmetrical potential well $U(r) = -A\delta(r-R)$. Neglecting the spin-orbital interaction, we will write the Hamilton operator of the system in the following form

$$\hat{H} = \hat{H}_0 + V, \qquad (1)$$

where

$$\hat{H}_0 = -\frac{1}{2}(\nabla_1^2 + \nabla_2^2) - A\delta(r_1 - R) - A\delta(r_2 - R) \qquad (2)$$

is the Hamilton operator of two free electrons in the potential well; $V = |\mathbf{r}_1 - \mathbf{r}_2|^{-1}$ is the operator of the Coulomb interaction between the electrons. Throughout the paper the atomic units (au) ($\hbar = m = |e| = 1$) are used. In the zero approximation (when the Coulomb interaction $V$ is neglected) the problem for the both electrons reduces to that in Refs. [4,5] considering the behavior of one electron in the Dirac-bubble potential. In this approximation each of the electrons with energy $E_s = -\beta^2/2$ obeys the wave equation

$$\left[-\frac{1}{2}\nabla_{1,2}^2 - A\delta(r_{1,2} - R)\right]\psi(\mathbf{r}_{1,2}) = -\frac{\beta}{2}\psi(\mathbf{r}_{1,2}). \qquad (3)$$

* According to [4], the extra electron is localized in the ground state on the $p$-like level. So the electron configuration $2p^z$ is quite acceptably. If one considers the ion ground state as a $s$-like state [5], then the state $1s^2 2p^{z-2}$ will correspond to minimal energy. It is evident that the absolute value of energy of this configuration is less than the upper estimate.

Here the strength of the delta-potential is defined by the formula [5,8]

$$A = \frac{\beta}{2}(1 + \coth \beta R). \tag{4}$$

Thus, in the Dirac-bubble potential model two experimentally observed parameters, namely the fullerene radius $R$ and the energy of the ground $s$-state $-E_s$, fully define equation (3) and hence the whole spectrum of electronic states of the negative ion $C_{60}$. The solutions of the wave equations for one electron with zero orbital moment $\psi_{1s}(\mathbf{r}_{1,2}) = [\chi_{\beta 0}(r_{1,2})/r_{1,2}]Y_{00}(\mathbf{r}_{1,2})$ can be represented as follows

$$\chi_{\beta 0}(r_{1,2}) = B\frac{\exp(-\beta R)}{\beta R}\sinh \beta r_{1,2} \qquad \text{for } r_{1,2} \le R,$$

$$\chi_{\beta 0}(r_{1,2}) = B\frac{\sinh \beta R}{\beta R}\exp(-\beta r_{1,2}) \qquad \text{for } r_{1,2} \ge R, \tag{5}$$

where $B$ is the normalized factor

$$B = \frac{z\sqrt{\beta}\exp(z/2)}{(\sinh z + \cosh z - z - 1)^{1/2}}; \quad z = 2\beta R. \tag{6}$$

In the first approximation of the perturbation theory the energy of the ground state of the system is equal to $E = E_0 + Q$, where $E_0 = 2E_s = -\beta^2$, and $Q$ is the mean value of the energy of the Coulomb interaction of two electrons in the potential well

$$Q(\beta) = \iint \Phi_\beta^*(\mathbf{r}_1,\mathbf{r}_2)\frac{1}{|\mathbf{r}_1 - \mathbf{r}_2|}\Phi_\beta(\mathbf{r}_1,\mathbf{r}_2)d\mathbf{r}_1 d\mathbf{r}_2. \tag{7}$$

Here $\Phi_\beta(\mathbf{r}_1,\mathbf{r}_2) = \psi_{1s}(\mathbf{r}_1)\psi_{1s}(\mathbf{r}_2)$ is the wave function of two non-interacting electrons; the integration in (7) is made over the six-dimensional configuration space. When integrating (7) the Coulomb potential is represented as a series in spherical function $Y_{lm}(\mathbf{r}_{1,2})$. The wave functions $\Phi_\beta(\mathbf{r}_1,\mathbf{r}_2)$ are independent of angular variables. Therefore, for integration over spherical angles of vectors $\mathbf{r}_{1,2}$, in this series all the terms except those with $l = m = 0$ will be zero. Thus, the integral (7) is transformed into the form

$$Q(\beta) = \int_0^\infty \chi_{\beta 0}^2(r_1)W(r_1)dr_1 \tag{8}$$

where

$$W(r_1) = \left[\frac{1}{r_1}\int_0^{r_1}\chi_{\beta 0}^2(r)dr + \int_{r_1}^\infty \frac{\chi_{\beta 0}^2(r)}{r}dr\right] \tag{9}$$

is the mean potential created by the electron cloud $\psi_{1s}(\mathbf{r})$ at the point $\mathbf{r}_1$. For $R$=3.527 Å= 6.665 au [9] the numerical value of the energy of Coulomb interaction $Q(\beta)$ in the well with the binding energy of one electron $E_s = -2.65$ eV is: $Q(\beta) = 3.72$ eV (In Sec. 1 this energy was estimated as 4.1 eV) while the energy of non-interacting electrons is: $E_0 = -2.65 \times 2 = -5.30$ eV. Hence the total energy of the system $E$ is negative.

The Dirac-bubble potential is a pseudopotential; its substitution in the wave equation gives a specified jump of the logarithmic derivatives $\Delta L$ of the wave function at the point $r$=$R$. In turn, the value $\Delta L$ is defined by the affinity energy of $C_{60}$ (for details see Refs. [4,5]). Formally this potential can be considered as that with zero-thickness $\Delta = 0$ and infinity depth. This potential is a limit case of the Lorentz-bubble potential

$$U(r) = -U_0 \frac{1}{\pi} \frac{d}{(r-R)^2 + d^2} \qquad (10)$$

for $d \to 0$. The maximal depth of the potential (10) at $r=R$ is $U_{\max} = U_0 / \pi d$. The thickness of the potential well $\Delta$ at the middle of the maximal depth is $\Delta = 2d$. The parameters of this potential $U_{\max}$ and $\Delta$ should be connected with each other in such a way that in the potential wells (10) there is a *s*-like state with the specified energy $E_s = -2.65$ eV. The parameters $U_{\max}$ and $\Delta$ are defined by solving the wave equation for electron moving in the potential well (10).

The calculation results for the wave functions $\chi_{\beta 0}(r)/r$ (*s*-like ground state) are presented in Fig. 1. For comparison, the wave functions calculated with the Dirac-bubble potential ($\Delta = 0$) are given in the same figures. With the rise in the parameter $\Delta$ the cusp-behavior of the wave function for zero-thickness changes to more smoothly behavior of wave functions near the point $r \approx R$. According to Fig. 1, the shape of the wave functions depends comparatively weakly on the shape of the potential wells in which the electron is localized.

Let us use these wave functions to calculate with the formula (7) the energies of the Coulomb interaction $Q(\beta)$ of electrons. The calculation results for different parameters of the potential wells are given in Table 1.

Table 1.

| $\Delta$, au | $U_{\max}$, au | $Q$, eV | $E$, eV |
|---|---|---|---|
| 0 | $\infty$ | 3.72 | -1.58 |
| 1 | 0.4415 | 3.68 | -1.62 |
| 2 | 0.2805 | 3.65 | -1.65 |
| 3 | 0.2243 | 3.62 | -1.68 |

A decrease in the energy of the Coulomb interaction $Q$ with the rise in $\Delta$ is quite explainable. The regions of electron delocalization increase and therefore the overlapping integrals of the wave functions (8) and (9) decrease. The total energy of electrons $E = Q - \beta^2$, locked in all the potential wells (10) is negative, so the existence of the double-charged negative ion of $C_{60}$ within the Lorentz- or Dirac-bubble potential models is quite acceptable.

**3. Variational method**

More exact values of energy and a wave function of the ground state of the double-charged negative ion $C_{60}$ can be obtained by a direct variational method. We replace the wave vector $\beta$ by $\gamma$ in the wave function $\Phi_\beta(\mathbf{r}_1, \mathbf{r}_2) = \psi_{1s}(\mathbf{r}_1)\psi_{1s}(\mathbf{r}_2)$ and in the one-electron functions $\psi_{1s}(\mathbf{r}_{1,2})$ and consider the wave vector $\gamma$ as a variational parameter. The problem of finding the energy of the ground state of two electrons in the Dirac-bubble potential well reduces to calculating the following integral

$$E(\gamma) = \iint \Phi_\gamma^*(\mathbf{r}_1, \mathbf{r}_2) \hat{H} \Phi_\gamma(\mathbf{r}_1, \mathbf{r}_2) d\mathbf{r}_1 d\mathbf{r}_2, \qquad (11)$$

and finding a minimum of the function $E(\gamma)$. The Hamilton operator of the system in (11) is defined by formulas (1) and (2). The integral (11) is divided into three parts: kinetic energy of electrons, potential energy of their interaction with the $C_{60}$ shell and electrostatic energy of electron interaction. As a result we obtain the following expression for the total energy of the system as a function of the wave vector $\gamma$

$$E(\gamma) = -\gamma^2 + 2[A(\gamma) - A] \,|\, \chi_{\gamma 0}(R) \,|^2 + Q(\gamma) \qquad (12)$$

where $Q(\gamma)$ is defined by formulas (8) and (9); $A(\gamma)$ is the parameter

$$A(\gamma) = \frac{\gamma}{2}(1 + \coth \gamma R) . \quad (13)$$

Another parameter $A$ is defined by the formula (4) and its numerical value is: $A \approx 0.44265$. Until now we impose no limitations on the vector $\gamma$. We define a value of this vector at the point where $dE / d\gamma = 0$. The minimum of the function $E(\gamma)$ corresponds to the ground state energy of the system in which the electrons are described by the trial wave function $\Phi_\gamma(\mathbf{r}_1, \mathbf{r}_2)$ being a product of the one-electron wave functions (5). The function $E(\gamma)$ is given in Fig. 2. The total energy of the system, according to Fig. 2, reaches minimum at $\gamma_{\min} \approx 0.4269$ and the ground state energy of the system is $E(\gamma_{\min}) \approx -1.589\,\text{eV}$. The first potential of the $C_{60}^{2-}$ double-charged ion detachment is $J_1 = -1.589 + 2.65 = 1.061\,\text{eV}$; the second one is $J_2 = 2.65\,\text{eV}$.

Thus, assuming that the trial wave function of the pair of electrons has the form of a product of the one-electron wave functions, we found the electronic structure of the $C_{60}^{2-}$ anion by the variational method. It is known that the energy found by the variational equation for the ground state of the system cannot be less than an exact value. That is if the Dirac-bubble potential model is correct then the double charged negative ion exists in a stable state and the electronic level of the ground state in the real anion is located deeper than $J_1 = 1.061\,\text{eV}$.

**4. $C_{60}^{2-}$ ion photodetachment**

More detailed information on the electronic structure of anions can be obtained by methods of photoelectron spectroscopy. Let us calculate the photodetachment cross sections of the $C_{60}^{2-}$ anion. Due to Coulomb interaction of electrons the threshold behavior of the photodetachment cross section changes significantly. Because of the Coulomb repulsion of electrons the cross section of the reaction $C_{60}^{2-} + \omega = C_{60}^{-} + e$ exponentially vanishes for $(\omega - J_1) \to 0$. While the photodetachment cross section of the one-charged ion near the threshold of the reaction $C_{60}^{-} + \omega = C_{60} + e$, according to the Wigner law, is proportional to $(\omega - J_2)^{3/2}$ [6].

Let us consider the first reaction as a result of which the double-charged ion $C_{60}^{2-}$ is transformed into a negative one-charged ion $C_{60}^{-}$. The radial part of the wave function of optical electron in the initial state is defined by the function $\chi_{\beta 0}(r)$ Eq.(5) in which the wave vector is $\beta = \gamma_{\min}$; the potential of the $s$-level splitting is equal to $J_1$. We will calculate the photodetachment cross section in the “frozen core” approximation. In other words, the continuum wave functions will be calculated in the undistorted field of $C_{60}^{-}$. The potential of this field is a sum of the bubble potential $U(r)$ and the potential (8) created by the charge of the electron residing in the potential well: $U(r) + W(r)$. The mean potential $W(r)$ is calculated with the wave function $\chi_{\gamma_{\min} 0}(r)$. In this approximation we neglect changes in the $C_{60}^{-}$ ion field caused by the photoelectron emission. The potential $U(r) + W(r)$ in which the optical electron moves after photon absorption is given in Fig. 3. In this figure the potential $U(r)$ is represented by Lorentzian (10) with the parameters $U_{\max} \approx 0.8996$ and $\Delta \approx 0.2$. These parameters, as those in Table 1, make it possible for the $s$-like ground state with the binding energy $E_s = -2.65\,\text{eV}$ to exist in the well $U(r)$.

The radial parts of the continuum wave functions $\chi_{kl}(r)$ with specified orbital moment $l$ obey the following wave equation

$$\frac{1}{2}\left[\frac{d^2\chi_{kl}}{dr^2}-\frac{l(l+1)}{r^2}\chi_{kl}+k^2\right]-[U(r)+W(r)]\chi_{kl}=0\,. \qquad (14)$$

Here $k^2/2=\omega-J_1$ is the kinetic energy of photoelectron. For large distances the potential $[U(r)+W(r)]_{r\to\infty}\to 1/r$ coincides with the Coulomb repulsion potential. Therefore, the wave functions $\chi_{kl}(r)$ at $kr>>1$ have the following asymptotic form

$$\chi_{kl}(r)\approx\sin\left[kr+\frac{Z}{k}\ln 2kr-\frac{\pi l}{2}+\delta_l\right]. \qquad (15)$$

Here the charge of electron cloud formed by the residing electron is equal to $Z=-1$. The photo-process cross section is defined by the formula

$$\sigma(\omega)=\frac{8\pi N_{10}}{3k}\alpha\omega\,|\langle k,1|d|10\rangle|^2\,. \qquad (16)$$

Here $\alpha$ is the fine structure constant; $N_{10}=2$ is the number of electrons in the potential well. The dipole matrix element in (16) is defined by the following integral

$$\langle k,1|d|10\rangle=\int\chi_{k1}r\chi_{\gamma_{\min}0}dr\,. \qquad (17)$$

The calculation result of the photodetachment cross section $C_{60}^{2-}+\gamma=C_{60}^{-}+e$ is given in Fig. 4. The $C_{60}^{-}+\omega=C_{60}+e$ reaction cross section is given in the same figure. In the latter case the potential $W(r)$ in the equation (14) is equal to zero; the radial part of the electron wave function in the initial state is defined by the function $\chi_{\beta 0}(r)$ Eq. (5); in the asymptotic (15) the charge Z is $Z$=0; the number of electrons is $N_{10}$ = 1. Besides the photodetachment cross sections of $C_{60}$ negative ions, the cross section of the atomic negative ion photodetachment calculated within the zero-range potential model [10] is given in the same figure. Dashed and dotted curves in Fig. 4 coincide with the curves for cross sections calculated in paper [4]. According to this figure, the disintegration cross sections of the fullerene anions $C_{60}^{z-}$ with $z$=1 and 2 have the values of the same order as the photodetachment cross sections of atomic ions with the same affinity energy.

**5. Conclusions**

It should be noted that the energy of the ground state of the $C_{60}^{2-}$ system calculated with the help of the variational equation cannot be less than the experimental value. The variational method as such cannot provide the location of the system ground level deeper than the experimental one. The essence of the approximation used in the paper was to select a trial wave function of the pair of electrons as a product of the one-electron wave functions. It was shown that already in this approximation the total energy of the system is negative. The next approximations can be obtained if instead of the simple product one can take a general expression for the wave function with several variational parameters appropriate for the problem in question. The next-order corrections can only increase the depth of level location corresponding to the ground state of the system. Consequently, within the Dirac-bubble potential model the $C_{60}^{2-}$ anion is stable and the detachment potential of this anion is equal to (or even greater than) the value calculated in this paper. We hope that the data presented herein will prompt experimental works to look into the matter, thereby promoting such developments.

**Acknowledgments**

The author is very grateful to Dr. I. Bitenskiy for useful comments. This work was supported by the Uzbek Foundation Award Ф2-ФА-Ф164.

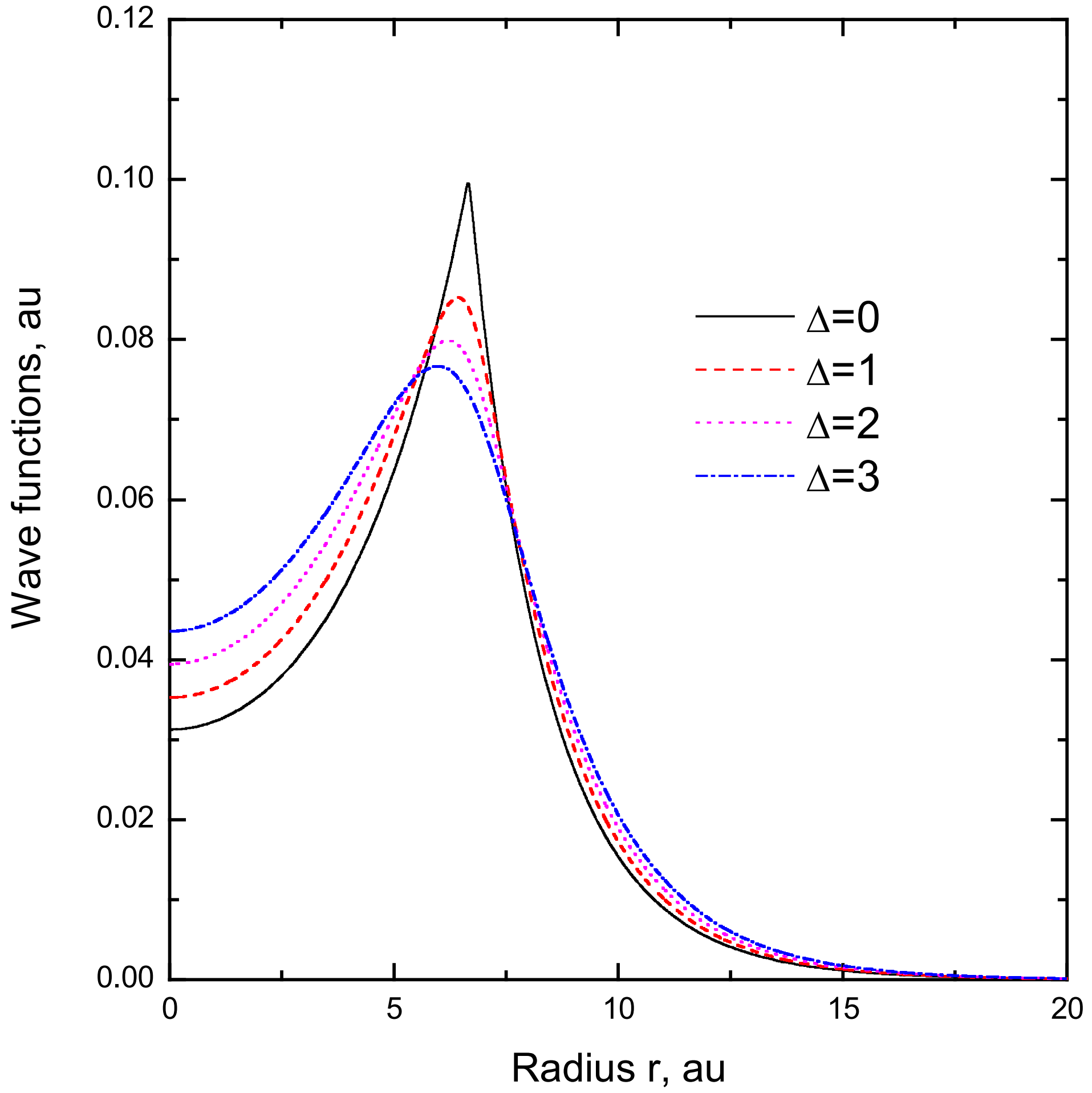

Fig. 1. The wave functions $\chi_{\beta 0}(r)/r$ with the different thicknesses $\Delta$ of the Lorentz-bubble potential well

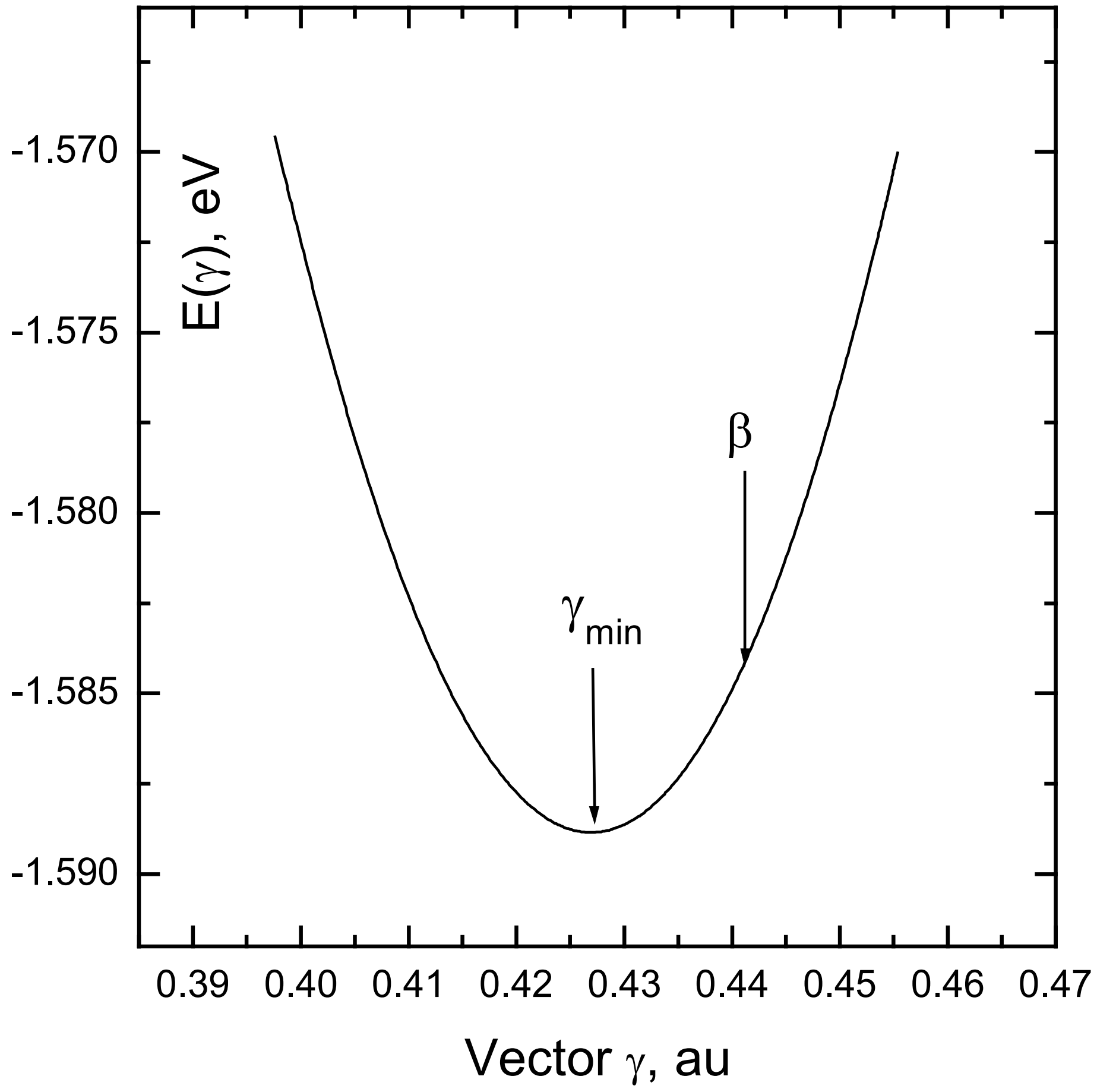

Fig. 2. The function $E(\gamma)$.

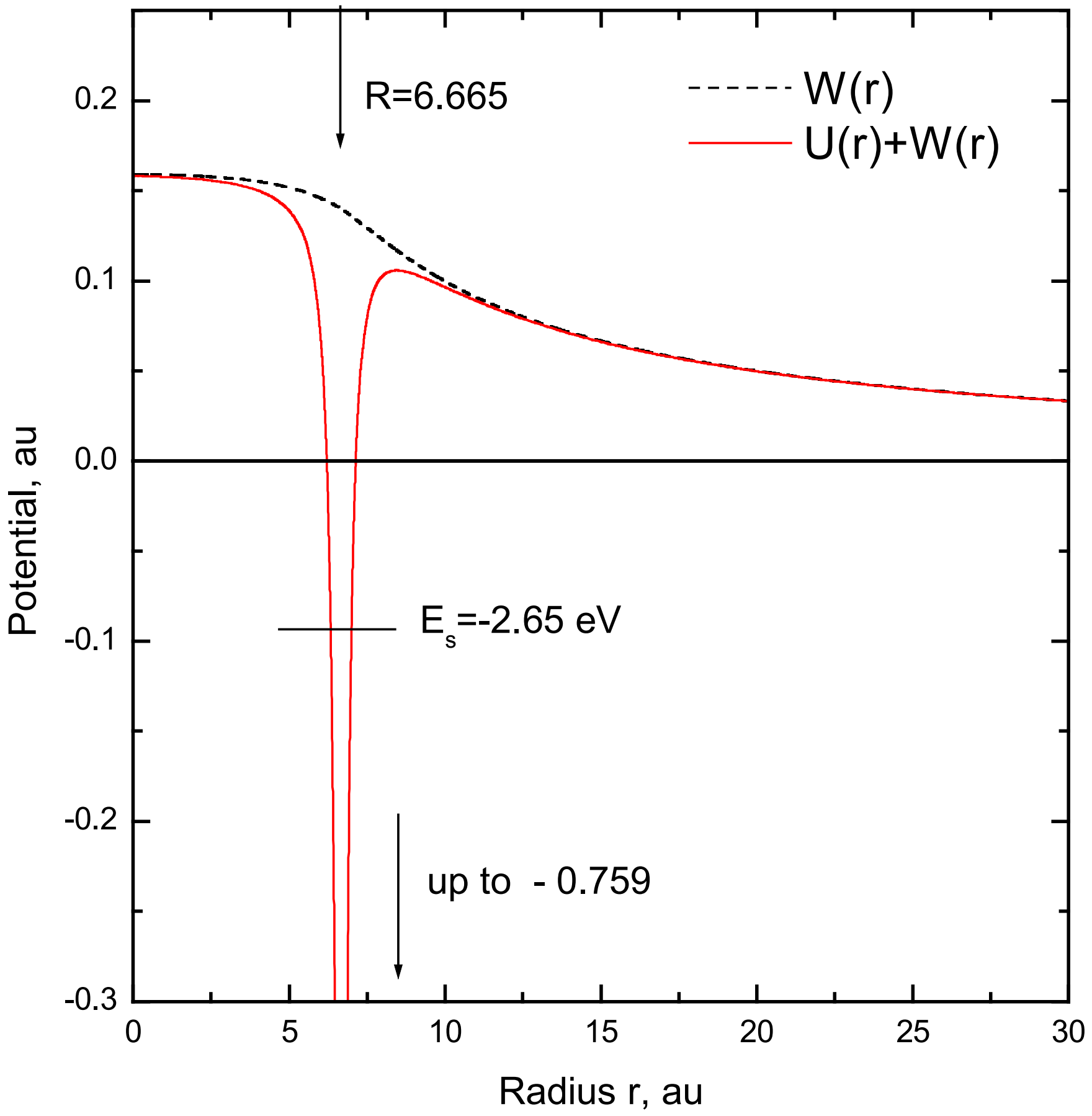

Fig. 3. The potentials $U(r)$ and $U(r)+W(r)$.

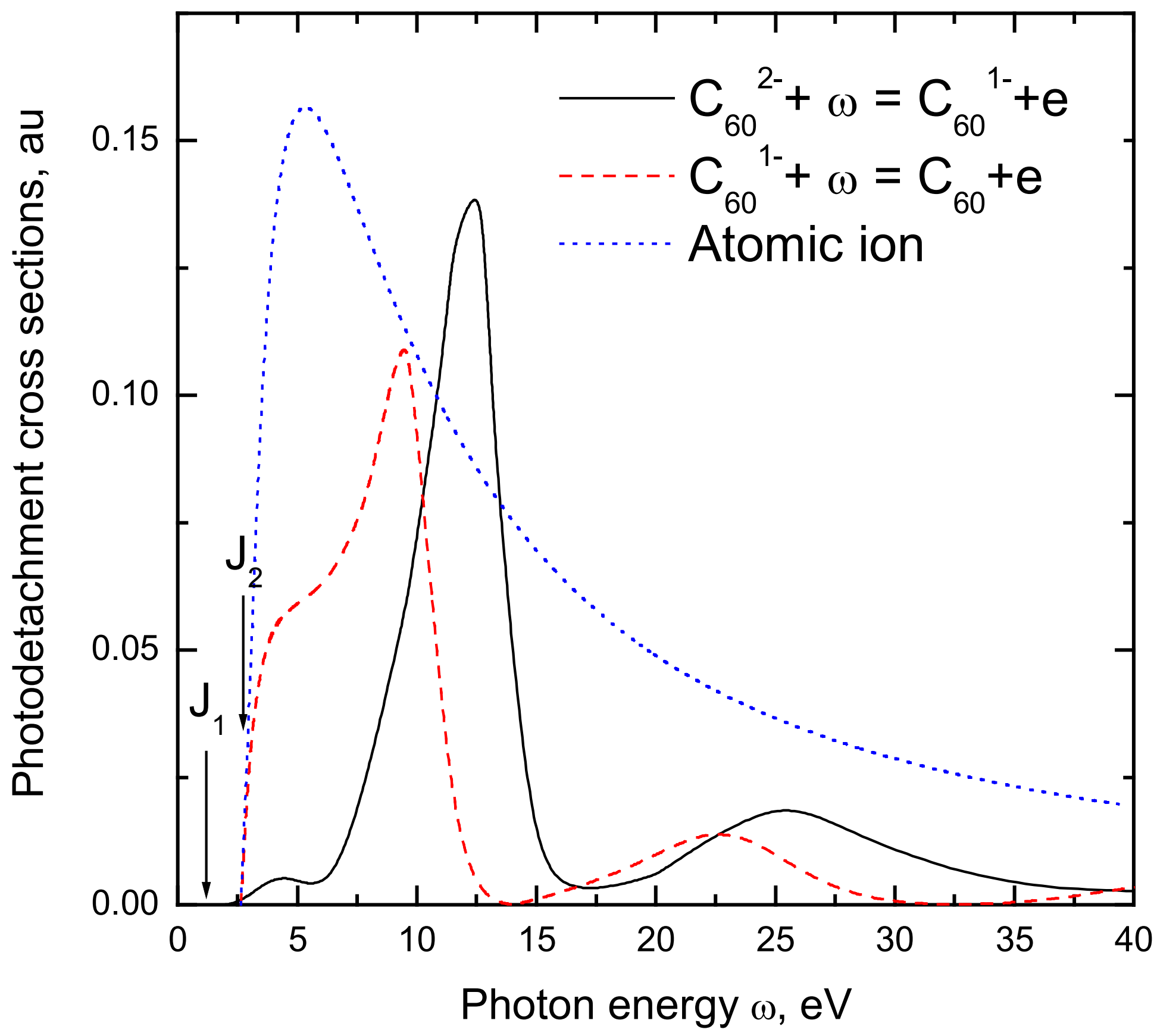

Fig. 4. Photodetachment cross sections.